\documentclass[twocolumn,english,aps,prl,noshowpacs,floatfix,superscriptaddress]{revtex4}
\usepackage[T1]{fontenc}
\usepackage[latin9]{inputenc}
\usepackage{amsmath}
\usepackage{amssymb}
\usepackage{graphicx}
\usepackage{float}
\usepackage{amsfonts}
\usepackage{physics}
\usepackage{xcolor}
\usepackage{algorithm}
\usepackage{algpseudocode}
\usepackage{url}
\usepackage{array}
\newcolumntype{P}[1]{>{\centering\arraybackslash}p{#1}}
\newcolumntype{M}[1]{>{\centering\arraybackslash}m{#1}}

\makeatletter
\@ifundefined{textcolor}{}
{%
 \definecolor{BLACK}{gray}{0}
 \definecolor{WHITE}{gray}{1}
 \definecolor{RED}{rgb}{1,0,0}
 \definecolor{GREEN}{rgb}{0,1,0}
 \definecolor{BLUE}{rgb}{0,0,1}
 \definecolor{CYAN}{cmyk}{1,0,0,0}
 \definecolor{MAGENTA}{cmyk}{0,1,0,0}
 \definecolor{YELLOW}{cmyk}{0,0,1,0}
}

\newcommand*{\balancecolsandclearpage}{%
  \close@column@grid
  \clearpage
  \twocolumngrid
}

\@ifundefined{definecolor}
 {\usepackage{color}}{}
\usepackage{siunitx}

\renewcommand{\fnum@algorithm}{\fname@algorithm}

\makeatother

\usepackage{babel}
\begin{document}
\flushbottom

\title{Boost clustering with Gaussian Boson Sampling: a full quantum approach}

\author{Nicol\`o~Bonaldi}

\affiliation{Data Reply s.r.l., Corso Francia, 110, 10143 Turin, ITALY}

\author{Martina~Rossi}

\affiliation{Data Reply s.r.l., Corso Francia, 110, 10143 Turin, ITALY}

\author{Daniele~Mattioli}

\affiliation{Enel S.P.A.}

\author{Michele~Grapulin}

\affiliation{Enel S.P.A.}

\author{Blanca~Silva~Fern\'andez}

\affiliation{Data Reply s.r.l., Corso Francia, 110, 10143 Turin, ITALY}

\author{Davide~Caputo}

\affiliation{Data Reply s.r.l., Corso Francia, 110, 10143 Turin, ITALY}

\author{Marco~Magagnini}

\affiliation{Data Reply s.r.l., Corso Francia, 110, 10143 Turin, ITALY}

\author{Arianna~Osti}

\affiliation{Enel S.P.A.}

\author{Fabio~Veronese}

\affiliation{Enel S.P.A.}


\begin{abstract}
	\section*{Abstract}
Gaussian Boson Sampling (GBS) is a recently developed paradigm of quantum computing consisting of sending a Gaussian state through a linear interferometer and then counting the number of photons in each output mode. When the system encodes a symmetric matrix, GBS can be viewed as a tool to sample subgraphs: the most sampled are those with a large number of perfect matchings, and thus are the densest ones. This property has been the foundation of the novel clustering approach we propose in this work, called \textit{GBS-based clustering}, which relies solely on GBS, without the need of classical algorithms. The GBS-based clustering has been tested on several datasets and benchmarked with two well-known classical clustering algorithms. Results obtained by using a GBS simulator show that on average our approach outperforms the two classical algorithms in two out of the three chosen metrics, proposing itself as a viable full-quantum clustering option.

\end{abstract}

\maketitle

\section{Introduction}
\label{Introduction}
Clustering is one of the most common unsupervised learning problems \cite{russell_artificial_2020}. Given a dataset, the goal is to group objects that are similar to each other with the purpose of having subsets that are meaningful in the specific context \cite{jain_algorithms_1988}. Therefore, a key concept is the definition of similarity, that must be tailored to each problem. This information can be provided as dissimilarity or design matrices whose values depend on the type of features that describe the input data \cite{hastie_elements_2009, murphy_machine_2012}. 

There are many clustering algorithms, the choice depends on the type and distribution of data \cite{xu_comprehensive_2015}. Among these, k-means clustering is widely used since it can be implemented relatively easily and is suitable for large datasets; however, it has some limitations such as convex cluster shapes that can be detrimental when data distributions don't match this structure \cite{macqueen_methods_1967}. In this case, an interesting approach is the DBSCAN, which is a density-based clustering algorithm. This method evaluates two parameters that determine if a region is dense and points in the same dense region are clustered together \cite{ester_density-based_nodate}. Nevertheless, DBSCAN is not appropriate for every data distribution; e.g. points of regions with different densities cannot be correctly clustered.

Considering advantages and disadvantages of classical algorithms, this work explores the potential benefits that a quantum approach to the clustering problem can achieve.

\medskip

Quantum computing promises to perform certain types of calculation considerably faster than classical computing, by exploiting quantum mechanics effects such as superposition, interference and entanglement \cite{nielsen_quantum_2011}. There exist several quantum computing paradigms, the main ones being \textit{quantum annealing} and \textit{universal quantum computing}. The first one \cite{kadowaki_quantum_1998} is mainly suitable for solving optimization problems formulated as Quadratic Unconstrained Binary Optimization (QUBO) problems \cite{lucas_ising_2014} and relies on the adiabatic quantum theorem \cite{albash_adiabatic_2018}. Once a system of qubits has been prepared depending on the model to be solved, it is left free to evolve towards the ground state, under the condition it corresponds to the optimal solution of the quadratic problem. On the contrary, universal quantum computing \cite{deutsch_quantum_1985} is based on a direct manipulation of qubits through quantum gates. This precise control of the quantum system guarantees a larger range of possible applications, but at the cost of requiring higher quality qubits \cite{lim2015experimental}.

\medskip

In 2011 Aaronson and Arkhipov \cite{aaronson_computational_2011} enriched the set of quantum computing protocols by introducing \textit{Boson Sampling} (BS), a novel model of quantum computation consisting of simultaneously sending identical photons through a linear interferometer and observing the output pattern in the photon number basis. They demonstrated that, under reasonable assumptions, BS is able to solve sampling problems which are beyond the capabilities of classical computing -in particular, sampling according to the permanent of a submatrix of the interferometer unitary- paving the way to the so called \textit{quantum advantage} \cite{terhal2018quantum,aaronson_computational_2011}. Although it does not demand a total control over the quantum system, a physical implementation of BS still needs perfectly deterministic sources of single photons, which happen to be extremely difficult to achieve. Even though there have been some experimental realizations of this protocol \cite{broome_photonic_2013, crespi_experimental_2013}, in order to ease the production of single input photons, several variants of Boson Sampling have been proposed (for instance, the Scattershot Boson Sampling \cite{lund_boson_2014}, which makes use of Gaussian states to improve the scaling of the generation probability of single photons which enter the interferometer). In 2017, Hamilton et al. \cite{hamilton_gaussian_2017} fully exploited the nature of Gaussian states by introducing a new protocol called \textit{Gaussian Boson Sampling} (GBS); contrary to other BS-inspired models, GBS's input consists of a Gaussian squeezed state, which guarantees significant experimental advantages. In 2022, Madsen et al. \cite{madsen_quantum_2022} implemented GBS on a photonic processor called Borealis and proved that this quantum protocol, when used to sample from a specific distribution, provides a significant computational advantage with respect to the best known algorithm running on a supercomputer. Moreover, several applications of GBS have been studied by Bromley et al. \cite{bromley_applications_2020}, ranging from graph similarity and graph optimization to molecular docking and quantum chemistry, showing that even though it is not a form of universal quantum computing, GBS offers a considerable versatility and can be used to efficiently solve several different problems. 

\medskip

In this work, we exploit the connection between Gaussian Boson Sampling and graph theory studied by Bradler et al. \cite{bradler_gaussian_2018} to develop a GBS-based clustering approach. The proposed quantum clustering technique aims to find clusters as dense regions of a properly constructed graph, relying solely on GBS, using a classical approach only in the post-processing phase to deal with isolated points. The approach has been tested on several datasets and benchmarked with the well-known k-means algorithm and DBSCAN, considering three different metrics. In the absence of a QPU which can encode any symmetric matrix, GBS has been performed by using a simulator provided by Xanadu \cite{noauthor_welcome_nodate}. Results show that our approach outperforms the two classical clustering algorithms when considering two of the three metrics and thus can be considered a viable full-quantum clustering option. 

\medskip

The rest of this paper is organized as follows: Section \ref{GBS} briefly introduces Gaussian Boson Sampling and its link to graph theory and the Hafnian of a matrix. Section \ref{GBS-based clustering} presents and explains the novel GBS-based clustering technique. The obtained results are collected in Section \ref{results} and deeply discussed in Section \ref{discussion}. Finally, conclusions and future works are presented in Section
\ref{conclusions}.

\section{Gaussian Boson Sampling}
\label{GBS}

Developed as an evoultion of Boson Sampling, Gaussian Boson Sampling is a specialized approach of photonic quantum computation, consisting of sending single-mode squeezed states into a linear interferometer. At the exit of the interferometer, detectors perform Fock state measurements on the obtained Gaussian state, counting the number of photons in each output mode.

As mentioned in the previous section, the clustering technique we propose in this work strongly relies on the connection between GBS and graph theory. Before introducing such a relationship, we start by reviewing the results about photo-counting from a Gaussian state presented in \cite{hamilton_gaussian_2017}. 

\medskip

Consider a system of $M$ \textit{qumodes}, namely $M$ optical modes of the quantized electromagnetic field. The state of this system can be univocally identified by a quasi-probability distribution described by the Wigner function, $W(\boldsymbol{p}, \boldsymbol{q})$, where $\boldsymbol{p} \in \mathbb{R}^M$ and $\boldsymbol{q} \in \mathbb{R}^M$ are called respectively the position and momentum quadrature operators. Gaussian states \cite{weedbrook_gaussian_2012} are those states whose Wigner function is a Gaussian distribution; as such, they are characterized by a $2M \cross 2M$ covariance matrix $\sigma$ and two $M$-dimensional vectors of means $\boldsymbol{\Bar{p}}, \boldsymbol{\Bar{q}}$. Now, let $\sigma_A$ be the covariance matrix of an arbitrary $M-$mode Gaussian state with zero mean and define the matrix $\mathcal{A}$ as:
\begin{equation}
\label{A_call}
    \mathcal{A} := X_{2M}[\mathbb{I}_{2M} - (\sigma_A + \mathbb{I}_{2M} / 2)^{-1}],
\end{equation}
where $\mathbb{I}_{2M}$ is the $2M-$dimensional identity matrix and \begin{math} X_{2M} := \begin{bmatrix}
0 & \mathbb{I}_{M} \\
\mathbb{I}_{M} & 0 
\end{bmatrix} \end{math}. 

Assume also that $\Bar{n}=\bigotimes_{i=1}^{M} n_i \ket{n_i}\bra{n_i} = (n_1, n_2, ..., n_M)$ corresponds to a specific output photon configuration, where $n_i$ is the number of photons measured in the $i$-th mode. Then, it can be shown \cite{hamilton_gaussian_2017, kruse_detailed_2019} that the probability of observing $\Bar{n}$ is
\begin{equation}
\label{prob_gbs}
    \mathbb{P}(\Bar{n}) = \dfrac{Haf(\mathcal{A}_{\Bar{n}})}{\Bar{n}! \sqrt{det(\sigma_Q)}},
\end{equation}
where $\Bar{n}! := n_1! n_2! ... n_M!$, $\sigma_Q := \sigma_A + \mathbb{I}_{2M}/2$ and $\mathcal{A}_{\Bar{n}}$ is a matrix associated to the observed output $\Bar{n}$. In particular, it is constructed starting from $\mathcal{A}$ as follows: if $n_i$ = 0, rows and columns $i$ and $i+M$ are removed from $\mathcal{A}$ and, if $n_i > 0$, rows and columns $i$ and $i+M$ are repeated $n_i$ times. Note that, when $n_i > 1$ for some $i$, this procedure produces a matrix which has no physical meaning (one can think of the repeated rows and columns to correspond to observed "pseudo-modes"); however, this allows one to link the $\mathbb{P}(\Bar{n})$ to the \textit{Hafnian} of a matrix in any output situation. The Hafnian of a $2M$-square matrix $B$ was introduced by Caianiello \cite{caianiello_quantum_1953} in the context of quantum field theory and is defined as 
\begin{equation}
\label{Hafnian}
    Haf(B) := \sum_{\mu \in PMP} \prod_{i=1}^M B_{\mu(2i-1), \mu(2i)},
\end{equation}
where $PMP$ is the set of perfect matching permutations.

\medskip

When sending states which have been squeezed according to a squeezing transformation $S$ through a linear interferometer described by a Haar random unitary $T$, the output Gaussian state has a covariance matrix $\sigma_A$ dependent on both $S$ and $T$ (see \cite{hamilton_gaussian_2017} for the explicit formula). Gaussian Boson Sampling has been introduced as the protocol which generates such a state and performs photo-counts measurements on it, according to Eq. (\ref{prob_gbs}).

\medskip

Returning to the photo-counts analysis, when the Gaussian state is \textit{pure}, the matrix $\mathcal{A}$ can be written as $\mathcal{A} = A \bigoplus A^*$, with $A$ an $M \cross M$ symmetric matrix, and the output probability distribution of the photo-counts becomes
\begin{equation}
\label{prob_gbs_pure}
    \mathbb{P}(\Bar{n}) = \dfrac{|Haf(A_{\Bar{n}})|^2}{\Bar{n}! \sqrt{det(\sigma_Q)}},
\end{equation}
where the matrix $A_{\Bar{n}}$ is constructed considering only rows and columns $i$ (and not $i,i+M$ as for $\mathcal{A}_{\Bar{n}}$ above). 

By relying on the above expression of $\mathcal{A}$, it is possible to efficiently encode any symmetric matrix $A$ into a GBS device \cite{bradler_gaussian_2018, bromley_applications_2020}. In other words, it is possible to set the squeezing transformation $S$ and the unitary $T$ in such a way that the produced Gaussian state has a covariance matrix $\sigma_A$ which guarantees that the Hafnian appearing in Eq. (\ref{prob_gbs_pure}) is computed on (possibly a submatrix of) a given symmetric matrix $A$. The proposed procedure exploits the Takagi-Autonne decomposition \cite{horn_matrix_1985} of $A$ and results in a pure Gaussian state. Equation (\ref{prob_gbs_pure}) becomes
\begin{equation}
    \label{prob_gbs_graph}
     \mathbb{P}(\Bar{n}) \propto c^s \dfrac{|Haf(A_{\Bar{n}})|^2}{\Bar{n}!},
\end{equation}
where $c$ is a rescaling parameter linked to the squeezing applied to the input modes and $s:=\sum_{i=1}^M n_i$. 

\medskip

Suppose now that the symmetric matrix $A$ encoded into the GBS machine is the adjacency matrix of an undirected graph $G$. As shown in \cite{barvinok_combinatorics_2016}, $Haf(A)$ corresponds to the number of \textit{perfect matchings} of $G$. A perfect matching of $G$ is a subset of edges of $G$ which match up every node of $G$ exactly once. Assessing the number of perfect matchings of a graph is a known difficult task for classical computers: in fact, it can be proven that this problem (which in turns corresponds to computing the Hafnian of the adjacency matrix) belongs to the $\#P-$complete complexity class \cite{valiant_complexity_1979}. However, thanks to the possibility of encoding any symmetric matrix into the GBS device, Gaussian Boson Sampling can be actually used to estimate the number of perfect matchings of an arbitrary graph $G$. In particular, it is related to the probability of observing $n=(1, 1, ..., 1)$, according to Eq. (\ref{prob_gbs_graph}). Note also that, if the output $n$ contains only 0s and 1s, it can be used to identify a subgraph of $G$ in the following way: if $n_i=1$, the $i-$th node of the graph is selected, whereas if $n_i=0$ the $i-$th node is discarded. In addition, Eq. (\ref{prob_gbs_graph}) states that the probability of observing a subgraph of the encoded graph $G$ is proportional to the square Hafnian of the corresponding adjacency matrix, so that subgraphs with a large Hafnian are sampled with a higher probability. In other words, a GBS machine can be prepared such that it samples, with high probability, subgraphs whose number of perfect matchings is large. Aaghabali et al. \cite{aaghabali_upper_2015} highlighted the connection between the number of perfect matchings in a graph and its density. In particular, the authors found a quantitative relationship between the two, confirming the intuition that a graph with a large number of perfect matchings is expected to contain many edges. Now the picture is complete: when sampling from a GBS device which encodes a graph $G$, the subgraphs that are most likely to appear are the dense ones. This fact has been exploited in \cite{arrazola_using_2018} to find dense subgraphs and is the foundation of our clustering algorithm.

\section{GBS-based clustering}
\label{GBS-based clustering}
Let $\{ x_i\}_i$ be a set of points to be clustered. Classical clustering algorithms such as k-means \cite{macqueen_methods_1967} group these points according to a distance function in such a way that close points belong to the same cluster. In particular, given the number of clusters $k$, k-means iteratively associates every element to the closest cluster and then it recomputes the cluster centers (``centroids''), until there are no more changes in the cluster composition. The nearest group is identified by computing the distance between data points and each cluster center; typically, Euclidean distance is used but any other distance metric can be implemented as well \cite{macqueen_methods_1967}. As mentioned before, k-means has some limitations linked to the clusters' shape: non-convex clusters which are not clearly separated are hardly identified. A different approach is adopted by DBSCAN, which is a density-based clustering method \cite{ester_density-based_nodate}. This algorithm uses two parameters, $\varepsilon$ and $MinPts$, to define clusters as dense regions. However, as mentioned in Section \ref{Introduction}, even DBSCAN is not suitable for every dataset. 

\medskip

\begin{figure}
   \centering
   \includegraphics[width=0.48\textwidth]{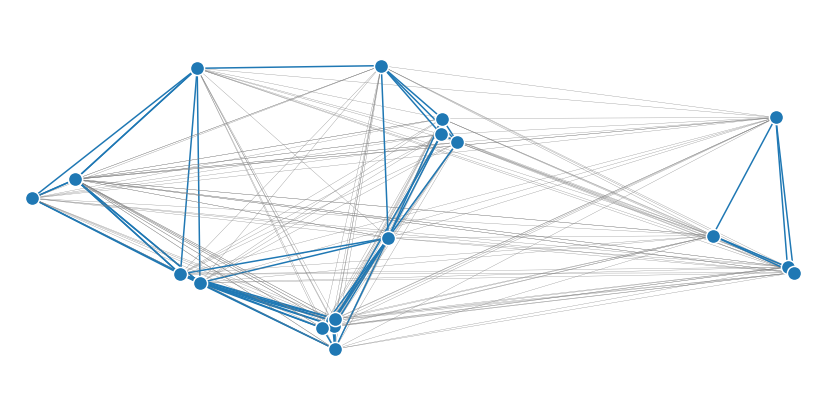}
   \caption{\textbf{Creation of the graph $G$ from the points $\{ x_i \}_i$}. Blue edges are shorter than the chosen threshold $\Tilde{d}$ and thus are selected. In the resulting graph, close points are connected and dense subgraphs can be considered clusters of points.
   }
   \label{fig: building graph}
\end{figure}

Our clustering approach, which we name \textit{GBS-based clustering}, adopts a different point of view. In Section \ref{GBS} we highlighted the relationship between GBS and graph theory. In particular, when sampling from the GBS distribution of a graph, the subgraphs that are most likely to appear are the ones with high density. Such subgraphs consist of points which are connected to each other and disconnected to points belonging to other dense subgraphs. If one thinks of a node of the graph as a point $x_i$ to be clustered and ensures that close points are connected, then dense subgraphs correspond to the common interpretation of clusters. Starting from this observation, the first step of our clustering approach (see Algorithm description at the end of this section) is to build a sparse graph $G$ from the points to be clustered. First, we compute the distance matrix $D$ of the $\{ x_i\}_i$ such that $D_{ij}:=d(x_i,x_j)$, $d$ being a distance. Then, we set a threshold $\Tilde{d}$ and we build the adjacency matrix $A$ which characterizes the graph $G$ as
\begin{equation}
\label{adjacency_mat}
     A_{ij} :=
    \begin{cases}
      1 & \text{if $D_{ij} < \Tilde{d}$} \\
      0        & \text{otherwise.}
    \end{cases}
\end{equation}
In other words, two data points $x_i$, $x_j$ are connected in $G$ if and only if their distance $D_{ij}$ is smaller than a chosen threshold $\Tilde{d}$. This way, we convert a list of points $\{ x_i\}_i$ into an undirected sparse graph $G$: by construction, dense subgraphs of $G$ consist of points which are close to each other and can therefore be considered clusters (see Figure \ref{fig: building graph}). 

\medskip

The algorithm is iterative and starts by considering the above graph $G$ and its adjacency matrix $A$. At each step, we obtain $N$ subgraphs by performing $N$ times GBS from the adjacency matrix $A$. We then identify the densest subgraph (in case of tie, the subgraph which has the largest number of nodes is considered) and if its density is higher than a threshold $t$, then it is chosen as a cluster, otherwise GBS is used to sample another set of $N$ subgraphs and the process is repeated. At each iteration, the threshold $t$ is lowered: this way, after few samplings, the probability of identifying a cluster is very high. Once the cluster is found, the corresponding nodes are discarded from the graph, the adjacency matrix $A$ is updated and the process restarts. This loop is performed until the number of nodes remaining in the graph is not too small. 

\medskip

\begin{figure}[H]
\centering
\begin{tabular}{cc}
     \includegraphics[width=.24\textwidth]{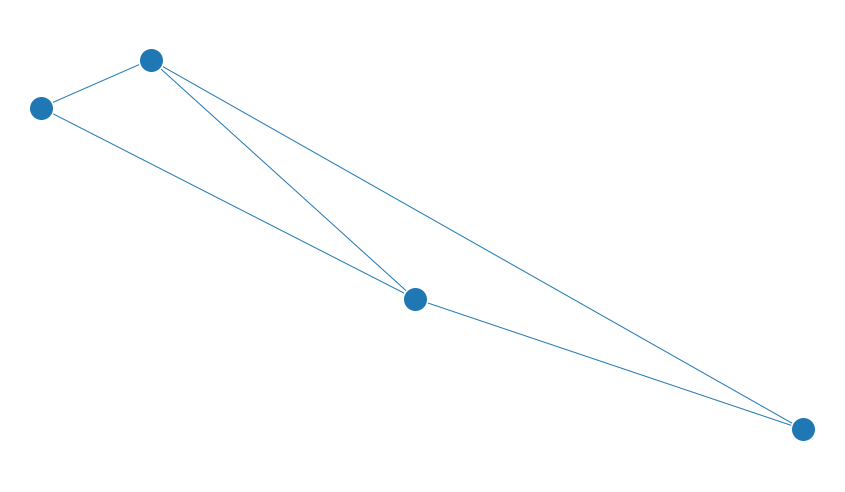} &
     \includegraphics[width=.24\textwidth]{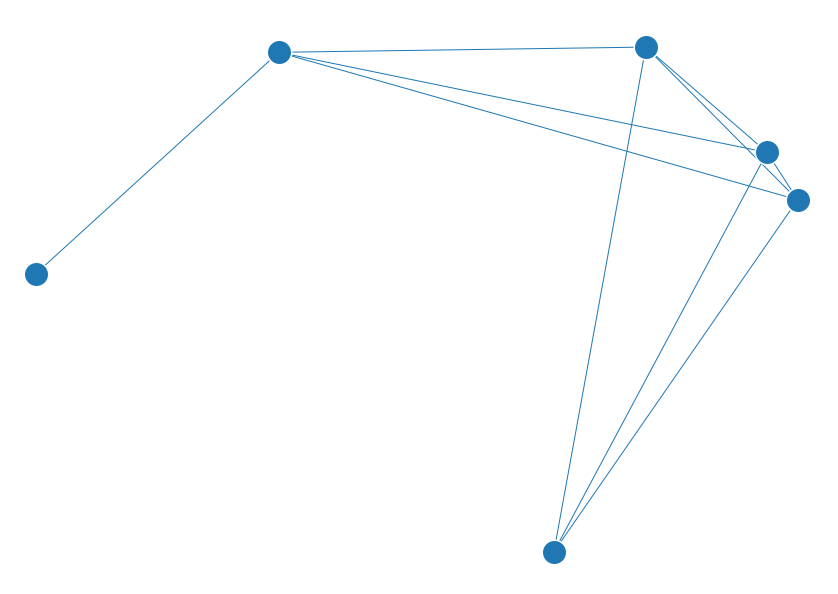} \\
    a) Small high-density graph  & b) Large low-density graph 
\end{tabular}
\caption{\textbf{Hafnian of $A$ vs density of $G$}. \textbf{a)} A small high-density graph: $\text{density}(G)=0.83$ and $Haf(A)=2$. \textbf{b)} A large low-density graph: $\text{density}(G)=0.66$ and $Haf(A)=3$. GBS samples the graph on the right with a larger probability, however it would not be a good cluster, since it is quite sparse.}
\label{fig: Hafnian vs density}
\end{figure}

Note that the check on the density of the subgraphs is crucial, mainly due to the fact that Eq. (\ref{prob_gbs_graph}) only guarantees that, when performing GBS, the more sampled subgraphs are those with a large Hafnian. However, a large low-density graph (namely one consisting of a large number of slightly connected nodes) could have a larger Hafnian than a small high-density subgraph (see Figure \ref{fig: Hafnian vs density}). In other words, the Hafnian of the adjacency matrix can be considered a reliable measure of the density of a graph only when comparing graphs which have the same number of nodes. 

Since the number of nodes composing a cluster is not known \textit{a priori}, the density check cannot be avoided. Following the same reasoning, among the $N$ sampled subgraphs, we post-select only those whose number of nodes is bigger than a threshold $L$. Indeed,  small graphs are more likely to be dense than larger graphs, since the number of possible edges in an $M$-nodes graph is $\mathcal{O}(M^2)$. However, in our framework, a cluster does not need to be extremely dense (as in the case of a \textit{clique} \cite{luce_method_1949}, i.e. a fully connected graph) because this would limit the quality of the clustering, by producing a large number of tiny clusters. On the contrary, it is solely required to own a certain degree of connection between nodes. Therefore the post-selection of samples is done to avoid that a very small subgraph which is highly dense is considered a cluster in place of a fairly less dense but larger subgraph. In other words, selecting only large subgraphs helps obtain \textit{maximal} clusters, namely clusters which cannot be enlarged preserving a high density.

\medskip

The process continues while the graph has a sufficient number of nodes and those which remain unclustered enter the post-processing phase. In this final step, each unclustered node $n$ is assigned to a cluster according to its connectivity. In particular, if $n$ is an isolated point, it forms a new cluster on its own; otherwise it is assigned to the cluster $c$ for which the ratio between the number of connections linking $n$ to $c$ and the number of nodes of $c$ is the highest.

\begin{algorithm}
\caption{GBS-based clustering}
\label{algorithm}
\begin{algorithmic}
\State $\Tilde{d}, n_{mean}, N, L \gets setParameters()$
\State $D \gets computeDistanceMatrix(\{ x_i \}_i)$
\State $A \gets buildAdjacencyMatrix(D, \Tilde{d})$
\State $clusters \gets EmpyList$
\While{$A$ is big enough}
    \State $i \gets 0$
    \State $Go \gets True$
    \While{$Go$}
    \State $s \gets GaussianBosonSampling(A, n_{mean}, N)$
    \State $s \gets postSelectSamples(s, L)$
    \State $best \gets findDensestCandidate(s, A)$
    \State $d_{best} \gets computeDensitySubgraph(best, A)$
    \State $t \gets computeThreshold(i)$
    \If{$d_{best} > t$}
        \State $clusters \gets clusters + best$
        \State $Go \gets False$
    \EndIf
    \State $i \gets i+1$
    \EndWhile
    \State $A \gets removeFoundCluster(A, best)$
    \EndWhile
\State $clusters \gets postProcessing(clusters, A, D)$
    

\end{algorithmic}
\end{algorithm}

\section{Results}
\label{results}

Since its development, Gaussian Boson Sampling has been implemented on several quantum hardwares by different research groups \cite{zhong_quantum_2020, madsen_quantum_2022}. However, when using the public available QPUs \cite{noauthor_borealis_nodate}, it is still not possible to encode a symmetric matrix to obtain a sample from a graph. For this reason, to test our clustering approach, we performed GBS by using a simulator provided by Xanadu \cite{noauthor_welcome_nodate}. This poses some limitations on the size of the symmetric matrix from which one can sample. In particular, we noticed that graphs with more than 30 nodes are extremely slow to sample from. For this reason, we tested the GBS-based clustering algorithm described in Section \ref{GBS-based clustering} on 30 datasets consisting of a variable number of locations $\{ x_i \}_{i=1}^M$, with $M$ ranging from 15 to 25, identified by their latitude and longitude. To guarantee a fair benchmark with k-means and DBSCAN, the usual Euclidean distance has been used to build the distance matrix $D$. In order to set a meaningful threshold $\Tilde{d}$ (which is then used to build the adjacency matrix of the graph), we tried different percentiles of the distribution of the distances appearing in matrix $D$; after a careful calibration on a large number of different datasets, we found that the best clusterings were obtained when setting $\Tilde{d} = D_{0.35}$, where $D_{0.35}$ is the $35^{th}$ percentile of $\{D_{ij}\}_{i,j}$. Gaussian Boson Sampling has been performed by using the \texttt{strawberryfields.sample.sample} function \cite{noauthor_sfappssamplesample_nodate}, which takes as input a symmetric matrix $A$, the mean number $n_{mean}$ of photons observed in output and the number $N$ of samples to produce. Some reasonable values were found to be $n_{mean} = size(A)/2$ and $N=50$; the parameter $L$, used to post-select large samples, has been set to $L=size(A)/3$. This choice should favor the creation of a large initial cluster but has just a negligible impact on following iterations.

\medskip

The clustering outcomes obtained with our approach have been compared to the results of k-means, where $k$ has been chosen for each dataset according to the so called \textit{elbow analysis}, and to the results of DBSCAN. Different values of the hyperparameters of the latter have been tested; here we report only the best results, obtained when using $\varepsilon = 0.005$ and $MinPts=2$. Additionally, noisy points found by DBSCAN have been clustered using the same post-processing function developed for our algorithm. To measure the quality of clustering, we used three metrics: the well-known \textit{silhouette score} \cite{rousseeuw_silhouettes_1987}, the weighted density of clusters $w$ and the intra-inter cluster cohesion $\delta_{ie}$. The first one relies on the Euclidean distance between points, whereas the other two exploit the graph structure built upon the data. In particular, $w$ is defined as $w:= \frac{\sum_i^k n_i \cdot d_i}{M} \in [0,1]$, where $d_i$ and $n_i$ are respectively the density and the cardinality of cluster $i$ and $k$ is the number of found clusters. Finally, $\delta_{ie} \in [-1, 1]$ is defined as the average difference $\delta_{int}-\delta_{ext}$, computed for each cluster. Given a cluster $i$, $n_i$ is the number of nodes in cluster $i$, $edge^{int}_i$ is the number of internal edges for the cluster and $edge^{ext}_i$ corresponds to the number of edges connecting cluster $i$ to any other point outside the cluster. Thus, $\delta_{int} := \frac{edge^{int}_i}{n_i(n_i - 1)/2}$ and $\delta_{ext} := \frac{edge^{ext}_i}{n_i(n - n_i)}$. For each metric, the higher the value, the better the clustering. In fact, a high silhouette score implies that, on average, a point is well paired with its assigned cluster. Concerning the weighted density $w$, it means that clusters are dense subgraphs, namely highly connected sets of points. Recall that, by construction, connectivity between points is strongly related to their proximity, therefore dense subgraphs correspond to high-quality clusters. Finally, a high value of intra-inter cluster cohesion $\delta_{ie}$ means that, not only the clusters have high density, but also that they are disconnected to each other and therefore points belonging to different clusters are far away. The mean results over the 30 datasets are reported in Table \ref{mean_metrics} and discussed in the following section.

\begin{table}[h!]
\centering
\begin{tabular}{P{0.22\linewidth} | P{0.22\linewidth} | P{0.22\linewidth} | P{0.22\linewidth} }
 \hline 
 \textbf{Method} & \textbf{Silhouette score} & \textbf{Weighted density} & \textbf{Intra-inter cluster cohesion}\\ [0.5ex] 
 \hline\hline
 k-means             & avg=0.40 std=0.06        & avg=0.65 std=0.08       & avg=0.52 std=0.12 \\ 
 \hline
 DBSCAN             & avg=0.29 std=0.11       & avg=0.73 std=0.11       & avg=0.71 std=0.11 \\
 \hline
 GBS-based clustering             & avg=0.33  std=0.10      & avg=0.83  std=0.09      & avg=0.77 std=0.09 \\[0.5ex]
 \hline
\end{tabular}
\caption{\textbf{Mean results and standard deviations over 30 datasets.} k-means produces the best silhouette score, but GBS-based clustering outperforms it when considering the weighted density $w$ and the intra-inter cluster cohesion $\delta_{ie}$. DBSCAN offers good results in terms of density and cohesion, at the cost of a poor silhouette score. Looking at the measured standard deviations, all of the three methods share similar variabilities and are stable with respect to different datasets.}
\label{mean_metrics}
\end{table}

\medskip

Finally, it is important to note that, when measuring the GBS output, two options can be experimentally realized: it is possible to count the exact number of photons in each mode or to use threshold detectors, which measure only the presence of photons in each mode. Evidently, the first way is more precise, since the output of the second method is composed only of 0s and 1s, which correspond respectively to "no photons" and "at least one photon". Equations (\ref{prob_gbs})-(\ref{prob_gbs_pure})-(\ref{prob_gbs_graph}) rely on the first method of measurement: when there is more than one photon (say $p$) in an output mode, the corresponding row and column of matrix $A$ are selected $p$ times. Thus, if A is the adjacency matrix of a graph $G$, the resulting sampled graph is no more a subgraph of the original $G$, but a new graph where some nodes have been repeated along with their connections. For this reason, the ideal way of measurement when performing the proposed GBS-based clustering would be to count the exact number of photons in each mode and post-select just those samples containing only 0s and 1s, which correspond to actual subgraphs of $G$. However, when using a simulator, counting the photons is an extremely slow operation: because of that, we decided to perform GBS using threshold detectors (setting \texttt{threshold=True} in the \texttt{sample} function). In this case, the exact probability distribution does not rely on the Hafnian of the matrix, but on its \textit{Torontonian}, a matrix function introduced and discussed in \cite{quesada_gaussian_2018}. The relationship between Hafnian, Torontonian and density of a graph has been studied in \cite{deng_solving_2023} through Monte Carlo simulations. The authors show a positive correlation between the Hafnian and the Torontonian and between the Torontonian and the density of the graph, validating the use of threshold detectors in GBS-measurement. Accordingly, we found that the proposed clustering method works properly even when using this faster approximate method of measurement. 

\begin{figure*}
\centering
\begin{tabular}{ c c }
     \includegraphics[width=.48\textwidth]{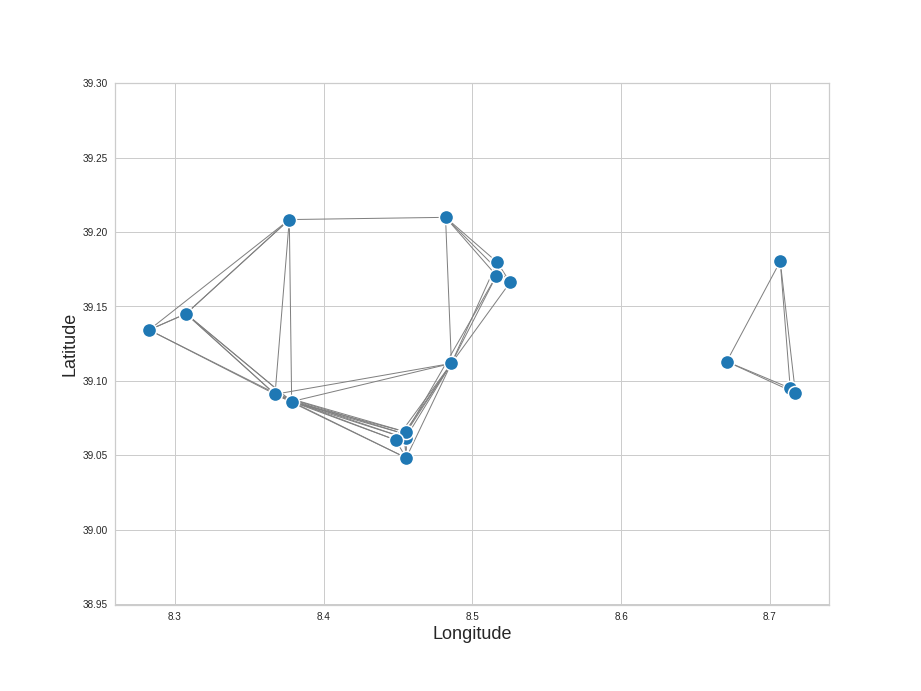} & \includegraphics[width=.48\textwidth]{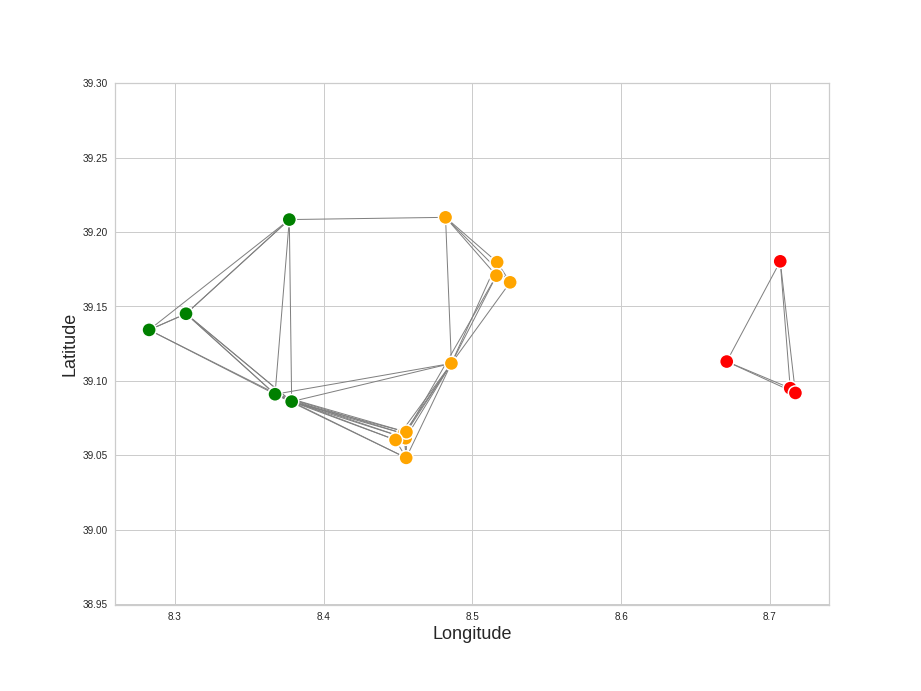} \\
    a) Unclustered points in graph $G$  & b) K-means \\
    \includegraphics[width=.48\textwidth]{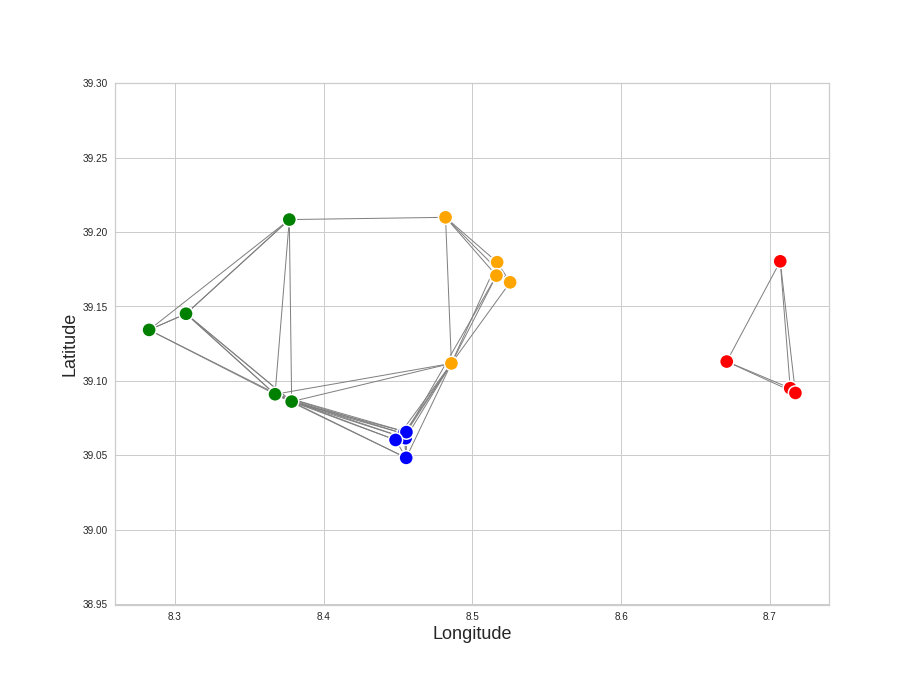} & \includegraphics[width=.48\textwidth]{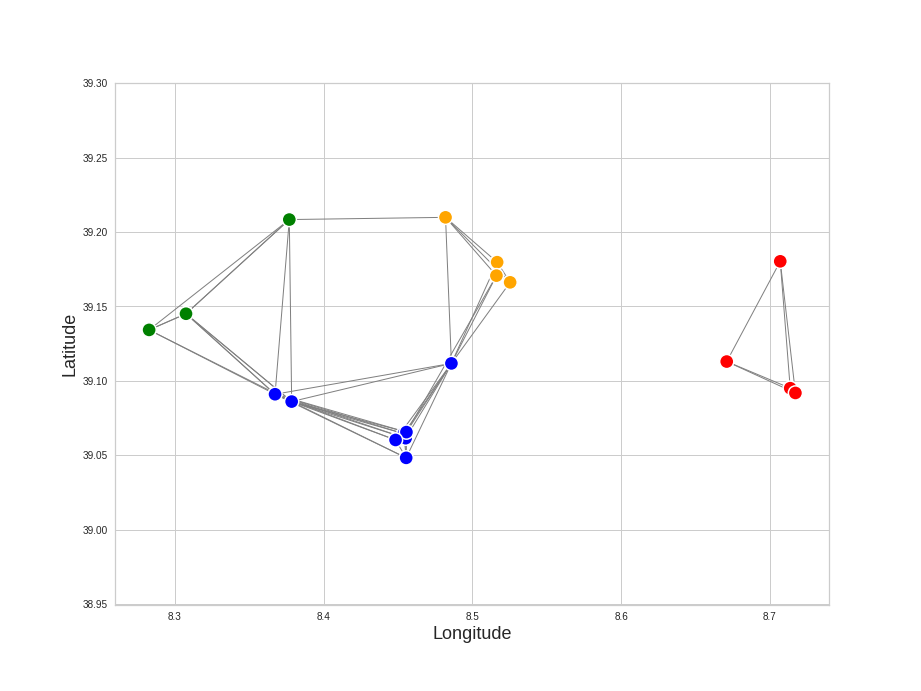} \\
    c) DBSCAN & d) GBS-based clustering
\end{tabular}
\caption{\textbf{Benchmark between different clustering methods on a selected dataset}. \textbf{a)} The points to be clustered embedded in the sparse graph $G$. \textbf{b)} Results obtained using k-means. The elbow analysis suggested $k=3$. \textbf{c)} Results obtained using DBSCAN. \textbf{d)} Results obtained with the GBS-based clustering. It is evident that the best clusterings have been obtained with the two methods which consider the density of points (DBSCAN and GBS-based clustering). This visual intuition is confirmed by every considered metric: $sil_{k-means}=0.54$, $sil_{DBSCAN}=0.61$, $sil_{GBS-based \ clustering}=0.61$;  $w_{k-means}=0.79$, $w_{DBSCAN}=1$, $w_{GBS-based \ clustering}=1$; $\delta_{ie, k-means}=0.76$, $\delta_{ie, DBSCAN}=0.87$, $\delta_{ie, GBS-based \ clustering}=0.91$. However, averaging over the 30 datasets, our approach performs better than
DBSCAN.}
\label{fig: benchmark}
\end{figure*}

\section{Discussion}
\label{discussion}
Every main clustering algorithm requires a choice of some parameters: for instance, in k-means it is the number $k$ of clusters; in DBSCAN, they are the $\varepsilon$, which defines the radius of the neighbourhood of a point, and $MinPts$, which is the minimum number of points of a cluster. In the proposed GBS-based clustering, $\Tilde{d}$ is the main parameter to set. It is responsible for the creation of the auxiliary graph $G$: large values produce a graph which contains a lot of edges, at the risk of connecting points which are not close; small values, instead, generate a high sparse graph, where fairly close points are not connected and therefore have low chance of being clustered together. Note that this crucial threshold has a concrete meaning, since it defines the \textit{vicinity} between points. Therefore, in real scenarios, one can leverage this fact and set $\Tilde{d}$ according to their definition of proximity. In this work, however, since the points to be clustered do not represent real-world datasets, we set $\Tilde{d}$ in order to obtain a number of clusters which was similar to the one obtained when using k-means. Other parameters to choose are $n_{mean}$ and $N$: the first one has been set quite large in order to favor the sampling of large subgraphs, whereas $N$ is allowed to be small, since Eq. (\ref{prob_gbs_graph}) guarantees that, with high probability, the GBS-device automatically samples graphs with a large number of perfect matchings. Finally, in our analysis, where the number of points to be clustered ranged between 15 and 20, the threshold $L$ has a tangible impact only at the first iteration of the algorithm, when the first cluster is identified. After an accurate analysis, we set $L=M/3$. Although the chosen thresholds performed as expected, a rigorous way of setting them could be investigated in a future work.

\medskip

Results shown in Table \ref{mean_metrics} demonstrate that despite the GBS-based clustering shows a smaller silhouette score than k-means, it performs much better when considering the other two metrics. On the contrary, DBSCAN produces a poor silhouette but good results in $w$ and $\delta_{ie}$. As it was mentioned before, this is expected, since it is a density approach. However, our algorithm is able to surpass DBSCAN in every considered metric. Moreover, in several datasets, GBS-based clustering happened to be the best approach, getting an even higher silhouette than k-means (for instance, see the dataset reported in Figure \ref{fig: benchmark}). Note that, even though the graph $G$ has no physical meaning and is constructed only to exploit the link between GBS and graph theory, clusters of points $\{ x_i \}_i$ should be dense with respect to $G$ anyway. Indeed, a dense cluster represents a set of points which are \textit{pairwise} close: the actual distance is neglected, but is guaranteed to be below a certain threshold $\Tilde{d}$. K-means produces clusters in a way that a point is assigned to the closest centroid, which is more loosely related to the points of its cluster: clusters don't have a high density. On the contrary, DBSCAN takes the density of points into account, but loses focus on the global picture, getting poor results in terms of the actual distance. Given the obtained results and also considering a certain intrinsic variability due to the random nature of GBS, we believe that the proposed GBS-based clustering is able to capture the density of clusters while not neglecting the effective distance between points, resulting in a method which is at least as good as k-means and DBSCAN. We should stress that, if quantum hardware were available, we could have counted the exact number of photons in each mode and post-selected samples corresponding to proper subgraphs of $G$. This precise method of measurement could have even improved the results.

\medskip

A remark on the scalability of the proposed approach. The number of possible outcomes of GBS scales exponentially, so that, as $M$ increases, the number of samples $N$ required to estimate the actual probability distribution of photon patterns becomes immediately huge. Nevertheless, in our approach, we are not interested in estimating $\mathbb{P}(\Bar{n})$ (nor some Hafnian). Conversely, we know from Eq. (\ref{prob_gbs_graph}) that the most sampled subgraphs are those with a large number of perfect matchings, independently of $N$ and $M$. For this reason, by using an accurate quantum hardware, we still expect to be able to produce dense subgraphs by setting a value of $N$ which guarantees that our GBS-based clustering remains computationally feasible.

\medskip

Finally, recent works such as \cite{oh_quantum-inspired_2023, solomons_gaussian-boson-sampling-enhanced_2023} have focused on the analysis of \textit{lossy} Gaussian Boson Sampling, namely one containing imperfections. By means of numerical simulations on graphs generated using the Erd\H{o}s--R\'enyi form, they show that, even in the presence of loss and spectral impurity, the samples obtained by GBS do not seem significantly different from the ones obtained when using a perfect GBS. If the results of these analysis were confirmed and extended to a general graph, there would be two consequences; first, the advantage of using GBS in finding dense subgraphs would likely be at most polynomial, since a lossy GBS can be efficiently simulated by classical algorithms. Second, GBS could be realized on a GBS device with few requirements in terms of loss and purity and thus our clustering algorithm could be implemented on a real quantum hardware in the short term. To confirm these results, more study is needed.

\section{Conclusions}
\label{conclusions}

Clustering is an unsupervised learning task which finds application in a plethora of real-world and research contexts. For this reason, it is crucial to develop clustering algorithms which can outperform well-known methods and quantum computing could be the key element to achieve this goal. 

\medskip

In this work, we propose an innovative clustering approach, which relies on Gaussian Boson Sampling (GBS), a recently developed model of quantum computation. This paradigm is strongly related to graph theory and can be used to sample high-density subgraphs from a parent graph. By exploiting this property of GBS, our algorithm identifies clusters of points as dense regions of a suitably constructed graph. 

\medskip

The proposed method has been tested on 30 datasets, using a GBS simulator which posed some limitations on the number of points that can be clustered. When a real quantum hardware able to encode any symmetric matrix is available, we expect GBS to be performed in a faster and more precise way. In particular, it will be possible to count the exact number of photons in each output mode, leading to the Hafnian-version of GBS (instead of the implemented Torontonian-version), which we believe could produce even more accurate results. Nevertheless, the obtained results demonstrate that, on average, our approach outperforms k-means and DBSCAN on two out of the three chosen metrics, proposing itself as a viable full-quantum clustering option. To further prove this point, a more complete benchmark between our method and other classical algorithms could be the subject of a future work.

\medskip

Finally, it is important to note that, in this work, a suitable graph is constructed starting from the points to be clustered. However, the same algorithm could be applied directly to a given graph, to solve what is called \textit{graph partitioning}, namely the task of finding communities in a network. In this perspective, future work will be focused on the case of weighted graphs and on the possibility of having overlapping clusters.

\section*{Acknowledgments}
\begin{acknowledgments}
This research work was supported by Enel S.P.A. that has funded the activity. 
\end{acknowledgments}

%


\bibliography{gbs_clustering}{}
\bibliographystyle{ieeetr}

\balancecolsandclearpage

\end{document}